\documentclass[aps,prl,reprint, floatfix]{revtex4-2}

\usepackage{import}
\usepackage{custom_aps}
\usepackage{siunitx}

\PassOptionsToPackage{full}{textcomp}	
\usepackage{newtxtext,newtxmath}		
\crefname{section}{Sec.}{Secs.}			
\crefname{figure}{Fig.}{Figs.}			
\Crefname{figure}{Figure}{Figures}		
\crefrangelabelformat{subfigure}{#3#1#4--#5(\crefstripprefix{#1}{#2}#6}	

\usepackage[caption=false]{subfig}		
\newcommand{\phantomsubfloat}[1]{
	\captionsetup[subfigure]{labelformat=empty}	
	\subfloat[][]{#1}}

\pdfoutput=1

\definecolor{linkcolour}{rgb}{0.02,0.12,0.3}
\hypersetup{colorlinks=true,citecolor=linkcolour,linkcolor=linkcolour,urlcolor=linkcolour}

\DeclareSIUnit{\gauss}{G}
\newcommand{\ideal}{\circ}

\begin{document}

\title{Shift of the Bose--Einstein condensation temperature due to dipolar interactions}

\author{Milan Krstaji\'{c}}	
\author{Ji\v{r}\'{i} Ku\v{c}era}
\author{Lucas R. Hofer}
\author{Gavin Lamb}
\author{Péter Juhász}
\author{Robert P. Smith}
\email{robert.smith@physics.ox.ac.uk}

\affiliation{Clarendon Laboratory, University of Oxford, Parks Road, Oxford, OX1 3PU, United Kingdom}

\date{\today}
	
\begin{abstract}
We report the first measurements of the BEC critical temperature shift due to dipolar interactions, employing samples of ultracold erbium atoms which feature
significant (magnetic) dipole-dipole interactions in addition to tuneable contact interactions. 
Using a highly prolate harmonic trapping potential, we observe a clear dependence of the critical temperature on the orientation of the dipoles relative to the trap axis. Our results are in good agreement with mean-field theory for a range of contact interaction strengths. This work opens the door for further investigations into beyond-mean-field effects and the finite-temperature phase diagram in the more strongly dipolar regime where supersolid and droplet states emerge. 

\end{abstract}

\maketitle

The phase transition to a Bose--Einstein condensate (BEC) is unique in that it is not driven by interactions but can occur solely due to quantum statistics. However, in the myriad of systems and phenomena where Bose condensation plays a critical role---from superconductivity, to liquid helium, to exciton and magnon condensation---inter-particle interactions are always present and often strong. The interplay of interactions and Bose condensation is thus an important one, and an obvious question concerns the effect of interactions on the critical point. Surprisingly, even for weak short-range interactions, this is a theoretically difficult question to answer due to non-perturbative correlations that develop near the critical point. The conclusion (after a 50-year-long theoretical debate) was that for a homogeneous system, repulsive contact interactions enhance condensation, i.e. lead to a positive shift of the transition temperature \cite{Andersen:2004, Arnold:2001, Baym:2001, Holzmann:2004, Kashurnikov:2001}.   

Ultracold gases provide an ideal setting in which to experimentally address such questions. However, the finite (and often inhomogeneous) traps used to confine these gases further complicate the picture. In the case of a harmonically trapped gas with repulsive contact interactions, there is a geometric mean-field effect which reduces the critical temperature $T_{\rm c}$ \cite{Giorgini:1996}. Measurements, utilising tuneable interactions in a potassium gas, showed agreement with this mean-field prediction for weak interactions as well as evidence of an upturn at stronger interactions due to beyond-mean-field effects \cite{Smith:2011}. 

The physics becomes even richer when going beyond just contact interactions. In particular, experiments on ultracold atoms that additionally experience dipole-dipole interactions (DDI) have uncovered new single-droplet \cite{Chomaz:2016}, droplet-array \cite{Kadau:2016, Ferrier-Barbut:2016}, and supersolid phases \cite{Bottcher:2019, Tanzi:2019, Chomaz:2019, Sohmen:2021}, alongside the usual BEC. However, these experiments mostly focused on the zero-temperature regime, and the effect of dipolar interactions on the transition temperature (to a BEC or more exotic phases) has not been measured.  

In this Letter, we report the first measurements of the BEC critical temperature shift due to dipolar interactions. Using an ultracold dipolar Bose gas of \ce{^166Er} atoms in a cigar-shaped harmonic trap we reveal how $T_{\rm c}$ depends on both the orientation of the dipoles (relative to the trap axis) and the relative strength of dipolar and contact interactions. Our results show a clear dipolar effect that is broadly in line with mean-field theoretical predictions \cite{Glaum:2007, Glaum:2007b} and also begin to explore the region where such predictions may no longer be valid.

\begin{figure}
    \phantomsubfloat{\label{fig:phases}}
	\phantomsubfloat{\label{fig:dipoles}}
	\centering
	\includegraphics[width=\textwidth]{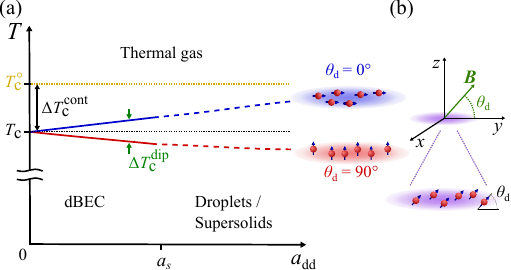}
	\caption{The dipolar $T_{\rm c}$ shift. \, (a) A cartoon phase diagram for a dipolar Bose gas showing the transition temperature, $T_{\rm c}$,  between the thermal gas and the condensed phase. Depending on the value of the dipolar length $a_{\mathrm{dd}}$, the degenerate gas will either be in the dipolar BEC state (dBEC), or for $a_{\mathrm{dd}} \gtrsim a_s$, it may collapse and form a dipolar droplet or supersolid phase. Repulsive contact interactions alone (dotted black line) cause $T_{\rm c}$ to reduce by $\Delta T_{\rm c}^{\textrm{cont}}$  compared to the ideal gas case ($T_{\rm c}^{\ideal}$, dotted yellow line). DDI are predicted to further modify $T_{\rm c}$ with the sign and magnitude depending on the dipole orientation. (b) The dipole orientation is set by the direction of the magnetic field~$\mathbf{B}$, that can be freely varied in the $y$--$z$ plane, changing the tilt angle~$\theta_{\rm d}$ w.r.t. the trap's long($y$)-axis.}
    \label{fig:fig1DipoleCartoon}
\end{figure}

\begin{figure*}
    \phantomsubfloat{\label{fig:cloudfits}}
    \phantomsubfloat{\label{fig:measNT}}
    \phantomsubfloat{\label{fig:TcPlot}}
    \subfloat{\includegraphics[width=0.5\textwidth]{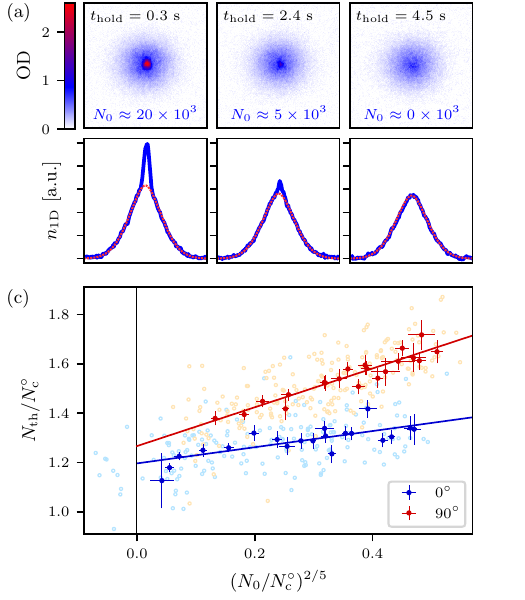}}
    \subfloat{\includegraphics[width=0.5\textwidth]{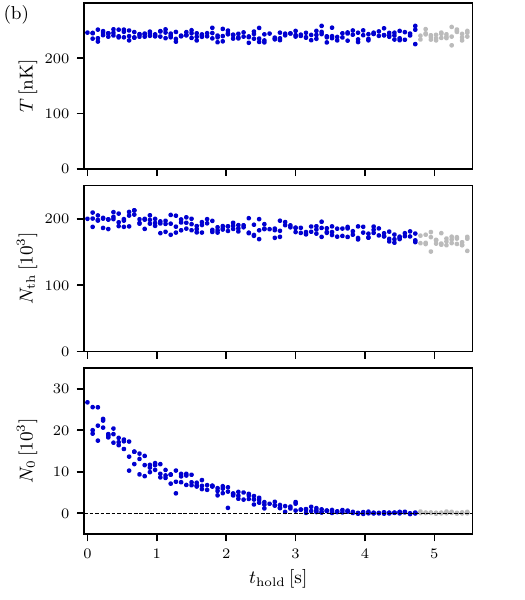}}
    \caption{Example of a critical point measurement. Here a harmonic trap with frequencies $\left(\omega_x, \omega_y, \omega_z\right) = 2\pi \times \left(294, 14.0, 233 \right)\, \mathrm{Hz}$ was used and $a_s=72 \, a_0$. (a) Absorption images and corresponding 1D sums of  $\theta_{\rm d} = 0^{\circ}$ clouds at three different hold times; the red lines are fits of the Bose distribution to the thermal component. 
    (b) Dependence of temperature, and thermal and condensate atom number (extracted from the absorption images) versus hold time for $\theta_{\rm d} = 0^{\circ}$.
    The data where we can confidently determine that no condensate is present (gray points) are filtered out before further analysis \cite{Supplement}. 
    (c) Dependence of $N_{\rm th}$ on $N_0$ for $\theta_{\rm d} \in \left\{ 0^{\circ}, 90^{\circ} \right\}$.  The hollow symbols show individual data points and the solid points represent binned data. The intercepts of the linear fits give the critical atom number, $N_{\rm c}$, and the gradients the magnitude, $S$, of the non-saturation effect. Note that small negative $N_0$ values [in both (b) and (c)] arise due to noise on our OD images \cite{Supplement, N025Footnote}.}
    \label{fig:TcExtract}
\end{figure*}

Figure 1 summarises the mean-field effect of both contact and dipolar interactions on the BEC critical point in an anisotropic harmonic trap; for simplicity we consider a cylindrically symmetric trap (with frequencies $\omega_{\parallel}$ and $\omega_{\perp}$). In this trap, an ideal gas with a fixed particle number $N$ has a critical temperature (yellow dotted line) given by $k_{\rm B} T_{\rm c}^{\ideal}=\hbar \bar{\omega} (N \bar{\omega}/\zeta(3))^{1/3}$ where 
$\bar{\omega}=(\omega_{\parallel}\omega_{\perp}^2)^{1/3}$ is the geometric mean of the trapping frequencies and 
$\zeta$ is the Riemann zeta function. The shift in $T_{\rm c}$, in the mean-field approximation, is given by \cite{Glaum:2007}:
\begin{equation} \label{eq:TcShiftCylTh}
    \frac{\Delta T_{\rm c}}{T_{\rm c}^\ideal} = -3.426  \frac{a_s}{\lambda} + 3.426  \frac{a_{\textrm{dd}}}{\lambda} f{\left(\frac{\omega_{\parallel}}{\omega_{\perp}} \right)} \frac{3 \cos^2{\theta_{\rm d}}-1}{4}.
\end{equation}

The first term on the r.h.s. arises due to contact interactions; this shift, $\Delta T_{\rm c}^{\rm cont}$, is negative and proportional to the ratio of the  $s$-wave scattering length $a_s$ and the thermal wavelength $\lambda = \sqrt{2 \pi \hbar^2/\left( m k_{\rm B} T_{\rm c}^{\ideal}\right)}$, where $m$ is the particle mass.

The second term describes the effect of DDI with strength characterised by the dipolar length $a_{\textrm{dd}} = \mu_0 \mu_{\rm d}^2 m/(12 \pi \hbar^2)$; here $\mu_{\rm d}$ is the magnetic moment of each atom. The anisotropic nature of the dipolar interactions means that the shift additionally depends on both the angle of the dipole orientation relative to the trap symmetry axis ($\theta_{\rm d}$) and the aspect ratio of the trap. The function $f{\left(\omega_{\parallel}/\omega_{\perp} \right)}$, defined in \cite{Glaum:2007} (see also \cite{Supplement}),  
is $\approx 1$ for the highly prolate `cigar' traps ($\omega_\parallel \ll \omega_\perp$) used in this work.

If one instead considers a system at fixed temperature, which is often more experimentally achievable, the ideal critical atom number is given by $N_{\rm c}^{\ideal} = \zeta(3)\left[k_{\rm B} T_{\rm c}^{\ideal}/\left(\hbar \bar{\omega}\right)\right]^3$ and any (small) shifts are related to \cref{eq:TcShiftCylTh} via 
\begin{equation} \label{eq:NcTcConversion}
   \frac{\Delta N_{\rm c}}{N_{\rm c}^{\ideal}} = -3 \frac{\Delta T_{\rm c}}{T_{\rm c}^{\ideal}}.  
\end{equation}
The fixed-temperature scenario also highlights a further important effect of interactions on Bose condensation, namely `non-saturation'. If $N$ is increased beyond $N_{\rm c}$, a condensate forms but the thermal component of the gas fails to saturate at $N_{\textrm{th}} = N_{\rm c}$ and instead continues growing with increasing condensate number $N_0$. This effect, which has been demonstrated for Bose gases with only contact interactions \cite{Tammuz:2011}, makes the measurement of the critical point more challenging, as the critical atom number $N_{\rm c}$ can no longer be identified with $N_{\textrm{th}}$ in a gas with (any) finite condensed fraction. Consequently, the $N_{\rm c} = N_{\textrm{th}}$ condition holds only at the transition point itself and one must therefore carefully extrapolate to $N_0 \rightarrow 0$. 

Our experiments use a gas of magnetically dipolar $^{166}$Er atoms, in the lowest Zeeman sublevel, for which $a_{\textrm{dd}}= 65.4\, a_0$ (with $a_0$ the Bohr radius) and $a_s$ can be tuned using a low-field ($\approx 0$\,G) Feshbach resonance~\cite{Patscheider:2022}; the relatively modest $a_{\textrm{dd}}$ is only expected to lead to dipolar effects on $T_{\rm c}$ at the few percent level, providing a significant experimental challenge. Our approach to this is to perform differential measurements whereby one seeks to directly measure the change in $T_{\rm c}$ due to changes in $\theta_{\rm d}$ and $a_s$ while keeping everything else the same.
This results in the cancellation of most first-order, common-mode systematic errors, e.g. on the atom number, trap frequencies \footnote{Note that due to the tensor polarisability of erbium, a change in the dipole orientation can lead to a change in the trap frequencies. To compensate, we carefully measured the effect allowing us to appropriately adjust the optical dipole trap laser power during dipole rotations. Our measurements are consistent with the theoretically predicted tensor part of the polarizability \cite{Becher:2018}.} and temperature.

We begin by preparing \cite{Krstajic:2023} a partially condensed cloud in our cigar-shaped optical dipole trap and with a magnetic field of $B =1.4\,$G ($a_s=72\, a_0$) applied perpendicular to the trap axis. We then rotate the magnetic field to set the desired $\theta_{\rm d}$ and subsequently ramp the magnetic field to its final value to set $a_s$. To vary $N$, we introduce a variable hold time  $t_{\textrm{hold}}$ during which the trap depth is chosen to ensure the temperature remains constant 
\footnote{Note that the hold time is either immediately before or after the $B$-field ramp.
Also note that the constant temperature comes about due to a balance between heating (e.g.\,due to inelastic collisions or 1-body trap heating) and residual evaporation.}. 
Finally, we turn off the trap and use absorption imaging on the broad $\SI{401}{\nano\metre}$ transition to record atomic density profiles after 18-24\,ms time-of-flight (ToF) expansion. For a given data run, we perform three interleaved sets of measurements: two at our chosen $a_s$ at both $\theta_{\rm d}=\ang{0}$ and \ang{90} and one calibration series with $a_s=72\, a_0$ and $\theta_{\rm d}=\ang{90}$. 
The calibration series effectively allows us to perform a differential measurement with respect to the same (fixed) point; see the supplemental material \cite{Supplement} for more details.

\Cref{fig:cloudfits} shows example images, and corresponding integrated 1D density profiles, for three different hold times for our data series with $a_s=72\, a_0$ and $\theta_{\rm d}=\ang{0}$.   
From the images we obtain the data triplet of $\left(T, N_{\mathrm{th}}, N_0\right)$ \cite{Supplement}. 
Note that the effect of interactions on the atom cloud during ToF (and thus the fitted temperature) depends on the dipole orientation (and $a_s$) \cite{Tang:2016}. So, as they are not common-mode, we correct for these effects; see \cite{Supplement}. 
In \cref{fig:measNT} we show how the three extracted quantities vary with $t_{\textrm{hold}}$.  
We see that the temperature remains essentially constant while the atom number slowly decays and $N_0$ decreases, eventually reaching zero.    

To determine the critical atom number, taking non-satuarion effects into account, we plot $N_{\textrm{th}}$ versus $N_0^{2/5}$; shown in \cref{fig:TcPlot} for $a_s=72\, a_0$ and $\theta_{\rm d}=\{\ang{0}$, \ang{90}\}. Plotting this way allows a linear extrapolation to $N_0\rightarrow0$ for a purely contact interacting gas~\cite{Tammuz:2011} and we have numerically shown that for $\varepsilon_{\rm dd}=a_{\rm dd}/a_s<~1$ this remains true in the presence dipolar interactions (see \cite{Supplement}). To take any small (<10\%) variations of $T$ into account, we scale both $N_{\rm th}$ and $N_{0}$ by $N_{\rm c}^{\ideal}{\left(T\right)}$. 
Even without fitting the data we see a clear effect of dipole orientation on the thermodynamics of the BEC critical point. 

Before performing a linear fit to the data in \cref{fig:TcPlot}, the effect of random measurement and fitting noise needs to be considered carefully due to several subtle effects. Firstly, the presence of random noise in estimating $N_0$ and $T$ skews the linear fit due to the strongly non-linear power-law scaling of $(N_0/N_{\rm c}^{\ideal})^{2/5}$. To mitigate this, we bin raw data points grouping them by $t_{\textrm{hold}}$ (see \cite{Supplement} for further details) to produce the solid points in \cref{fig:TcPlot}.  Secondly, as significant errors are present in both the $x$- and $y$-axis variables, we utilise orthogonal distance regression \cite{Boggs:1990} when performing the linear fits. These fits yield a slope $S$ and an intercept $N_{\rm c}/N_{\rm c}^{\ideal}$; the transition temperature shift is then [see \cref{eq:NcTcConversion}] given by $\Delta T_{\rm c}/T_{\rm c}^{\ideal}=-1/3 \times(N_{\rm c}/N_{\rm c}^{\ideal}-1)$. 

\begin{figure}
	\centering
	\parbox[c]{2 in}{\raggedleft\includegraphics[width=1.99 in]{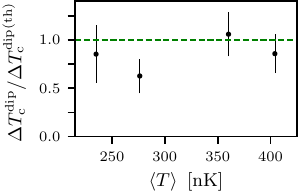}}
    \parbox[c]{1.2 in}{\raggedleft
    \begin{tabular}{c|c c c}
         $\left\langle T \right\rangle$ & $\omega_x$ & $\omega_y$ & $\omega_z$ \\
         $[\mathrm{nK}]$ & & $[\mathrm{Hz}]$ & \\
         \hline
         \footnotesize 235 & \footnotesize 294 & \footnotesize 14.0 & \footnotesize 233 \\
         \footnotesize 276 & \footnotesize 257 & \footnotesize 32.0 & \footnotesize 218 \\
         \footnotesize 360 & \footnotesize 360 & \footnotesize 26.9 & \footnotesize 269 \\
         \footnotesize 404 & \footnotesize 404 & \footnotesize 20.5 & \footnotesize 270 \\
    \end{tabular}
    \vspace{0.9 cm}}
    \caption{Dipolar $T_{\rm c}$ shift at $\varepsilon_{\rm dd}=0.91$ versus cloud temperature. The plot shows the ratio of $\Delta T_{\rm c}^{\rm dip}$ to the mean-field theoretical prediction; $\left\langle T \right\rangle$ is the mean temperature for each data series. The table shows the trapping frequencies used in each case.
    }
	\label{fig:TcvsT}
\end{figure}

To isolate the effect of the DDI on the transition temperature, evident in the different intercepts in \cref{fig:TcPlot}, we also calculate:
\begin{equation} 
    \label{eq:diffTcdip}
    \Delta T_{\rm c}^{\textrm{dip}} = \left. \Delta T_{\rm c} \right|_{\theta_{\rm d} = 0^{\circ}}-\left. \Delta T_{\rm c} \right|_{\theta_{\rm d} = 90^{\circ}}
\end{equation}
which for the data in \cref{fig:TcPlot} 
gives $\Delta T_{\rm c}^{\textrm{dip}}/T_{\rm c}^{\ideal}=2.3(0.9)\%$ in good agreement with the mean-field prediction of 2.7\%.
To confirm the robustness of this result at $a_s = 72\,a_0$ ($\varepsilon_{\rm dd}=0.91$) we performed similar measurements at three additional configurations with different cloud temperatures and trap shapes. This data is shown in \cref{fig:TcvsT} which confirms an overall good agreement with the mean-field prediction.

\begin{figure}
	\centering
	\phantomsubfloat{\label{fig:dTcvsas}}
    \phantomsubfloat{\label{fig:Svsas}}
    \parbox[c]{\columnwidth}{\raggedleft\includegraphics[width=\columnwidth]{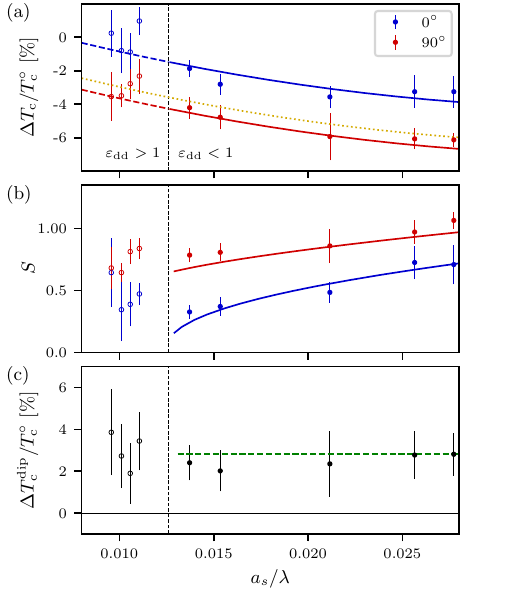}}
    \caption{
    Critical temperature shift in a dipolar gas. \,(a) $\Delta T_{\rm c}$, at different values of the scattering length $a_s$ and for $\theta_{\rm d} = 0^{\circ}$ and $\theta_{\rm d} = 90^{\circ}$; measurements performed in the same trap as in \cref{fig:TcExtract}. The lines represent the theoretical expectation (see text);
    solid blue for $\theta_{\rm d}=0^{\circ}$, solid red for $\theta_{\rm d}=90^{\circ}$ and dotted yellow for a non-dipolar gas [c.f.\@ \cref{fig:fig1DipoleCartoon}].
    (b) The measured slopes of the non-saturation effect compared with mean-field numerical predictions based on the Thomas-Fermi approximation (which only gives a stable solution for $\varepsilon_{\textrm{dd}}<1$).
    (c) The dipolar contribution to $\Delta T_{\rm c}$ which is simply the difference between the red and blue data points in (a).
    The green dashed line represents the mean-field theoretical prediction for $\Delta T_{\mathrm{c}}^{\mathrm{dip}}$. The open symbols denote points where $\varepsilon_{\rm dd}>1$; here the extrapolation procedure for extracting $N_{\rm c}$ (based on a fit of $N_{\rm th}$ vs $N_0^{2/5}$) may no longer be theoretically justified even though empirically it still describes the data well.
    }
	\label{fig:TcEffect}
\end{figure}

Finally, in \cref{fig:TcEffect} we explore the effect of varying $a_s$ on $T_{\rm c}$. \cref{fig:dTcvsas} shows $\Delta T_{\rm c}/T_{\rm c}^{\ideal}$ versus $a_s/\lambda$ for $\theta_{\rm d}=\ang{0}$ and \ang{90}
\footnote{Note that 
to extract $\Delta T_{\rm c}$ (the shift from the non-interacting limit) we assume that $\Delta T_{\rm c}$ of our calibration data is given by the first-order mean-field result (cf. \cref{eq:diffTcdip} and~\cite{Supplement}); we emphasize that any breakdown of this assumption would \emph{only} result in a global (vertical) shift of all the data in \cref{fig:dTcvsas}.}.
The solid lines are the dipolar shift predicted by \cref{eq:diffTcdip} together with the previously measured contact interaction shift~\cite{Smith:2011} (which includes a term quadratic in $a_s/\lambda$ as well as the linear term in \cref{eq:diffTcdip});  our data show close alignment with this prediction.

\Cref{fig:Svsas} displays the fitted slopes for the same data, which also show good agreement with our calculated non-saturation slopes (solid lines). 
The calculations employ the mean-field model from Ref.~\cite{Tammuz:2011} with dipolar interactions included (see \cite{Supplement} for details); note that they cannot be performed for $\varepsilon_{\rm dd}>1$, as in this regime the gas is unstable within the Thomas--Fermi approximation. 

Lastly, in \cref{fig:TcEffect}(c), we plot $\Delta T_{\rm c}^{\textrm{dip}}$ highlighting the clear shift of the BEC transition temperature due to dipolar interactions across a range of $a_s$ and the very good agreement with mean-field predictions (green dashed line) for $\varepsilon_{\rm dd}<1$~\cite{Glaum:2007, Glaum:2007b}. 

We now briefly consider the region $\varepsilon_{\rm dd}>1$ where the mean-field theory is expected to break down and where the ground state may not be a `standard' BEC but rather a droplet or supersolid state that is stabilised by beyond-mean-field effects. Here we limit ourselves to modest values of $\varepsilon_{\rm dd}$ to ensure that our measurements remain close to equilibrium (the 3-body loss rate grows sharply as $a_s$ is reduced \cite{Krstajic:2023}). In this `strongly dipolar' regime, the linear $N_{\rm th}$ versus $N_0^{2/5}$ trend is no longer expected to hold. However, we empirically find that it still describes the data well. We find that the $T_{\rm c}$ values obtained using this approach broadly follow the extrapolated mean-field  predictions.
Future investigations of the strongly dipolar regime will require a careful choice of trap parameters to ensure equilibrium conditions and an improved understanding of the consequences of the new ground states on the non-saturation effect.

In conclusion, we performed a high-precision study of the BEC transition temperature in an ultracold dipolar gas with a range of contact interaction strengths. Our results show a clear effect of dipolar interactions on the transition temperature. This is most clearly seen by directly comparing the $T_{\rm c}$ for clouds with the dipoles orientated parallel and perpendicular to the axis of the cigar shaped trap confining the atoms. The magnitude of the effect is in good agreement with mean-field predictions. In the future, it would be interesting to explore both the more strongly dipolar regime and also different trap geometries (particularly more homogeneous traps) where beyond-mean-field effects are expected to become important.

We thank Christoph Eigen and Zoran Hadzibabic for comments on the manuscript.
This work was supported by the UK EPSRC (grants no.\ EP/P009565/1 and EP/T019913/1). R.~P.~S.\ and P.~J.\ acknowledge support from the Royal Society, P.~J.\ acknowledges support from the Hungarian National Young Talents Scholarship, M.~K.\ from Trinity College, Cambridge, J.~K.\ from the Oxford Physics Endowment for Graduates~(OXPEG), L.~R.~H.\ from Christ Church College, Oxford,  and G.~L.\ from Wolfson College, Oxford.

\bibliography{robsrefs}

\begin{thebibliography}{32}
\bibitem [{N02()}]{N025Footnote}%
  \BibitemOpen
  \href@noop {} {}\bibinfo {note} {In \cref{fig:TcPlot}, to accomodate negative
  $N_0$ values we use
  $N_0^{2/5}=\mathrm{sgn}(N_0)~\times~|N_0|^{2/5}$.}\BibitemShut {Stop}%
\end{thebibliography}

\end{document}


\title{Supplementary Material: Shift of the Bose--Einstein condensation temperature due to dipolar interactions}

\author{Milan Krstaji\'{c}}	
\author{Ji\v{r}\'{i} Ku\v{c}era}
\author{Lucas Hofer}
\author{Gavin Lamb}
\author{Péter Juhász}	
\author{Robert P. Smith}
\email{robert.smith@physics.ox.ac.uk}

\affiliation{Clarendon Laboratory, University of Oxford, Parks Road, Oxford, OX1 3PU, United Kingdom}
\date{\today}
	
\begin{abstract}
\end{abstract}

\maketitle

\section{Data Analysis}
In this section we describe how we:
\begin{itemize}
    \item Extract the quantities ($T$, $N_{\rm th}$, $N_0$) from the time-of-flight absorption images of our atom clouds.
    \item Correct for the systematic effects of interactions on our extracted temperatures.
    \item Use our calibration series to perform a series-by-series absolute atom number calibration.
    \item Filter and bin the data in preparation for performing an orthogonal distance regression of $N_{\rm th}$ versus $N_0^{2/5}$.
\end{itemize}

\subsection{Image fitting procedure}
 
\begin{figure} 
    \phantomsubfloat{\label{sm:fig:Exp}}
    \phantomsubfloat{\label{sm:fig:fitNth}}
    \phantomsubfloat{\label{sm:fig:fitT}}
    \phantomsubfloat{\label{sm:fig:fitN0}}
    \phantomsubfloat{\label{sm:fig:sumN0}}
	\centering
	\parbox[c]{\textwidth}{\includegraphics[width=\textwidth]{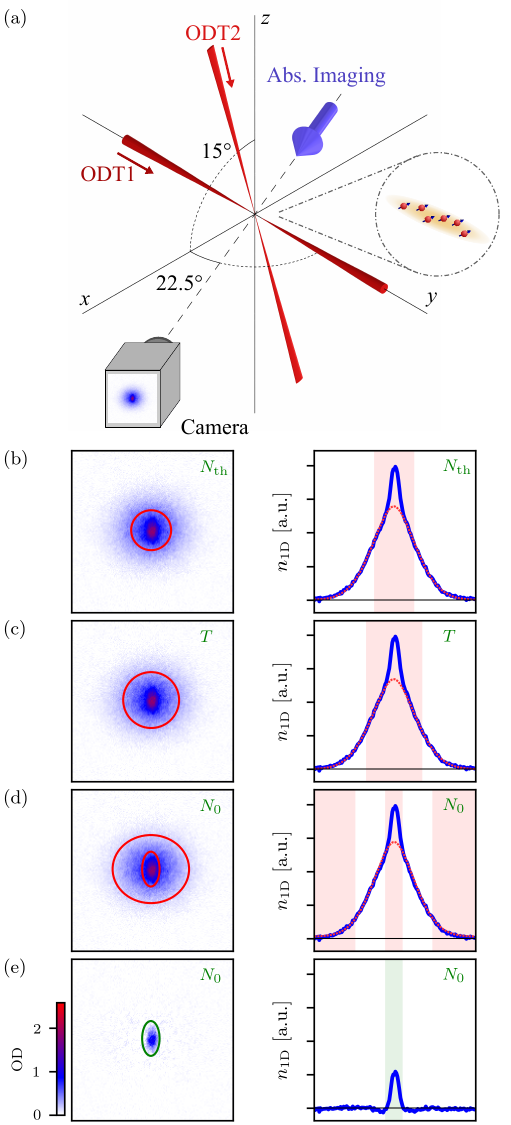}}
    \caption{Cloud imaging and fitting. \,(a) The layout of the experiment showing the optical dipole trap formed by two focused laser beams, ODT1 and ODT2. The imaging axis is in the horizontal plane at an angle of \ang{22.5} relative to the tight axis of our cigar shaped trap. (b-e)~Depict the procedure for estimating $N_{\textrm{th}}$, $T$ and $N_0$ as described in the text; left-hand plots show the OD images and the right-hand plots show the data integrated along the vertical axis. The red lines show the boundaries of the regions excluded from the 2D fits and the red shading illustrates which regions are excluded. In (e) the green ellipse shows the region which is summed to estimate $N_0$.}
    \label{sm:fig:fitting}
\end{figure}

The geometry of our imaging system is shown in \cref{sm:fig:Exp}; our absorption images display a signal that is proportional to the column density of atoms along the imaging direction.  We extract the thermal atom number, temperature and BEC atom number ($N_{\mathrm{th}}$, $T$, $N_0$) as described in \cref{sm:fig:fitting} and below. 

To extract $N_{\rm th}$ we apply a circular mask that is large enough to exclude the condensate for all our data [see \cref{sm:fig:fitNth}]. We then fit the thermal cloud with a Bose-enhanced column density distribution: 
\begin{equation}
    n_{\mathrm{th}}{(x,y)} = A g_2\left[e^{\frac{\mu}{\kB T}}
    e^{-\frac{1}{2} \left(\frac{x^2}{\sigma_x^2} + \frac{y^2}{\sigma_y^2}\right)}\right],
\end{equation}
where $g_2(z)$ is the the polylogarithmic function, $\mu$ the chemical potential, and $\sigma_x$, $\sigma_y$ are the cloud widths along the two principal axes. As we are concerned with partially condensed clouds, we set $\mu = 0$. The thermal atom number is then given by the integral of this fit function.

To extract $T$ we exclude a larger region from the centre of the cloud and rely only on the high-momentum tails of the distribution, which are least affected by (in-trap) interaction effects---see \cref{sm:fig:fitT}. The size of the excluded region [of around one thermal radius, $\sqrt{k_{\rm B}T/(m \omega^2)}$] was chosen to be large enough that the fitted temperature does not depend on the exact mask size but small enough to maintain a good signal to noise ratio. 
Assuming free expansion for a time of flight $t_{\rm ToF}$, the size of the cloud and the temperature are related by: 
\begin{equation}
\sigma_i^2=\frac{\kB T}{m \omega_i^2}+ \frac{\kB T t_{\rm ToF}^2}{m}.
\label{bec:eq:temp1}
\end{equation}
We determine the temperature from the vertical direction ($\sigma_z$) as this is both the least sensitive to the initial size of the thermal cloud (which can be modified due the presence of the BEC) and also the exact angle of our imaging beam with respect to the long axis of trap.

Finally, the condensate atom number $N_0$ is estimated by summing the residual density in the central part of the cloud once a fit to the thermal atoms has been subtracted.  For this purpose the fitting region [see \cref{sm:fig:fitN0}] excludes a tight region around the BEC and \emph{also} the wings of the distribution; this produces the best fit to the thermal cloud in the region close to the condensate. The size of both the central region excluded from the fit and the region used for summing [see \cref{sm:fig:sumN0}] depend on $N_0$ itself; for smaller condensates we use smaller regions, which reduces the absolute noise on $N_0$ to $\sim$\,$100$ atoms.

All the 2D fits are performed using a GPU-accelerated non-linear least-squares fitting procedure \cite{Hofer:2022}.  

\subsection{Temperature corrections}
\label{sec:Tcorr}

The presence of interactions can distort the cloud during its expansion, giving (small) corrections to the mapping of the cloud size to temperature. We consider, in turn, two independent effects, the self-interaction of the thermal cloud and the BEC-thermal interaction.

To address the self-interaction of the thermal atom cloud, we follow the approach of \cite{Tang:2016}. Here, the change of the momentum distribution of a harmonically trapped, non-degenerate, dipolar gas is calculated as a sum of a mean-field potential contribution (both contact and dipolar) and a hydrodynamic (collisional) contribution. For full details of the calculations, we refer the reader to Ref. \cite{Tang:2016} and its supplementary material. 
We note, however, that in order to account for the effects of Bose enhancement near criticality which were assumed negligible in~\cite{Tang:2016}, we replace all instances of the geometric mean of the trapping frequencies, $\bar{\omega}$, with an adjusted value $\bar{\omega}_a$. 

The role of this adjusted frequency is to approximate the Bose-enhanced density distribution with a Gaussian distribution (assumed in the model) which has a higher central density and is narrower than a standard Boltzmann distribution. Ensuring the peak phase-space density of a cloud with $N = N_{\rm c}^{\ideal}$ has the (correct) critical value, $\rho_{\rm c} = \zeta(3/2)$, requires that we choose
\begin{equation}
\label{eq:omega_correction}
\frac{\bar{\omega}_a}{\bar{\omega}}=\left(\frac{\zeta\left(3/2\right)}{\zeta(3)}\right)^{1/3} \approx 1.3 ~.
\end{equation}
Enforcing the correct central density is desirable given the dominant role of density (as opposed to the cloud shape) in both the mean-field effects and two-body collisions.
For our most strongly interacting clouds the temperature corrections from this effect are up to  5.5\% but the largest differential (between $\theta_{\rm d}=\ang{0}$ and \ang{90}) is never more than 0.5\% (excluding the points with $\varepsilon_{\textrm{dd}}>1$, where the differential may go up to 1.1\%).

To account for the BEC-thermal interactions and their effect on the atom distribution after ToF, we performed a numeric simulation of the cloud expansion in the presence of a mean-field potential produced by a dipolar BEC in the Thomas--Fermi (TF) regime (cf.\, l.h.s. of \cref{sm:eq:TFeq} with $V(\mathbf{r})=0$). In a single simulation run, \num{10000} test particles were initialized in a harmonic trap with positions and momenta sampled from the respective Gaussian distributions at an initial temperature $T_0$. For a given scattering length $a_s$, dipolar tilt angle $\theta_{\rm d}$, and BEC number $N_0$, the TF radii $R_{0,i}$ as a function of $t_{\rm ToF}$ are calculated (see below and \cite{Giovanazzi:2006}). The trajectories of the (thermal) particles in the time varying potential are numerically propagated using the velocity Verlet algorithm \cite{Verlet:1967} with forces given by the gradient of the effective potential. At the required ToF we numerically calculate the second moment of the simulated atom distribution, calculate an associated temperature and compare it to $T_0$. We find that the temperature corrections from this effect are linear in $N_0$ and constitute an up to  $\sim1\%$  correction on $T$; note that this effect vanishes for $N_0 \rightarrow 0$ so should mainly effect $S$ and not $T_{\rm c}$.

\subsection{Atom number calibration} 

Due to imperfect imaging we always observe an atom number $\Tilde{N} = \alpha N$ lower than the true value of $N$, where $\alpha <1$ is an unknown number which depends on e.g.\,imperfect polarization or frequency of the imaging beam. In this section we describe how we estimate $\alpha$ and more importantly how we make our results robust against any series-to-series variations in $\alpha$ or the exact trap frequencies.   

As described in the main text, our method for determining the transition temperature shift (and non-saturation slope) revolves around plotting $N_{\rm th}/N_{\rm c}^{\ideal}$ vs $(N_0/N_{\rm c}^{\ideal})^{2/5}$. This is based on the fact that we expect 
\begin{equation}
    \label{sm:eq:calib}
    \begin{split}
        \frac{N_{\mathrm{th}}}{N_{\rm c}^{\ideal}} = 
        1 + \frac{\Delta N_{\rm c}}{N_{\rm c}^{\ideal}} +
       S \left(\frac{N_0}{N_{\rm c}^{\ideal}}\right)^{2/5}
    \end{split}
\end{equation}
where $\Delta N_{\rm c}$ is the shift in the critical atom number from $N_{\rm c}^{\ideal} = \zeta(3)\left[ k_{\rm B} T_{\rm c}^{\ideal} / \left( \hbar \bar{\omega} \right) \right]^3$, the value for a harmonically trapped ideal Bose gas in the thermodynamic limit. 

If our measured atom numbers are related to the true atom numbers by $\Tilde{N} = \alpha N$ and our assumed trap frequencies are related to the true ones via $\Tilde{\omega} = \eta \omega$, we can rewrite \cref{sm:eq:calib} in terms of our measured (tilde decorated) quantities:
\begin{equation}
    \label{sm:eq:calib2}
    \begin{split}
        \frac{\Tilde{N}_{\mathrm{th}}}{\Tilde{N}_{\rm c}^{\ideal}} = 
        \beta \left(1 + \frac{\Delta N_{\rm c}}{N_{\rm c}^{\ideal}} \right)+
       \beta^{3/5} S \left(\frac{\Tilde{N}_0}{\Tilde{N}_{\rm c}^{\ideal}}\right)^{2/5}
    \end{split}
\end{equation}
where $\beta=\alpha \eta^3$. We see that to determine $\Delta N_{\rm c}/N_{\rm c}^{\ideal}$ and $S$ only requires that we know the single combined quantity $\beta$. 

We find $\beta$ for each data set using our calibration series (with $a_s = 72\, a_0$ and $\theta_{\rm d} = 90^{\circ}$) by assuming that at this point both $\Delta N_{\rm c}/N_{\rm c}^{\ideal}$ and $S$ are given by their theoretically expected values, and then performing a single parameter fit to find $\beta$.   

For our experiments,  $\Delta N_{\rm c}$ contains corrections due to non-interacting effects as well as the interaction shifts that we are seeking to measure:
\begin{equation}
     \Delta N_{\rm c} = \Delta N_{\rm c}^{\rm n.i.}+\Delta N_{\rm c}^{\rm int}.
\end{equation}
For the `theoretically expected' value  $\Delta N_{\rm c}^{\rm n.i.}$ we take into account contributions due to (i) finite atom number $\Delta N_{\rm c}^{\rm f.n.}$ and (ii) the combined effect of trap anharmonicity and finite trap depth $\Delta N_{\rm c}^{\rm a.h.}$.

\textbf{(i) Finite atom number correction}: Since the number of atoms in the cloud is not infinite, we need to account for the effect this has on the critical point (see e.g.\, \cite{Pethick:2002}):
\begin{equation}
    \label{sm:eq:fnshift}
    \frac{\Delta N_{\rm c}^{\mathrm{f.n.}}}{N_{\rm c}^{\ideal}} = 2.18\frac{\omega_m}{\bar{\omega}} (N_{\rm c}^{\ideal})^{-1/3}.
\end{equation}
Here, 
$\omega_m$ and $\bar{\omega}$ are the arithmetic and geometric mean of the trap frequencies, respectively. 

\textbf{(ii) Trap corrections}: As our trap potential $V(\textbf{r})$ is generated by two Gaussian laser beams in combination with gravitational field, it has a finite depth and also is not perfectly harmonic. 
For a gas of non-interacting bosons trapped in our potential, we can write the critical atom number as:
\begin{equation}
    \label{sm:eq:psdintegral}
    N_{\rm c}^{\rm trap} = \frac{1}{\left(2 \pi \hbar \right)^3} 
    \iint{
    \frac{1}
    {e^{\left[\epsilon\left(\textbf{r},\textbf{p}\right) \right]/ \left( \kB T \right)}-1} d\textbf{p} \, d\textbf{r}
    },
\end{equation}
where $\epsilon{\left(\textbf{r},\textbf{p}\right)} = \frac{p^2}{2m} + V\left(\textbf{r}\right)$ is the semi-classical energy of the system. Note that the upper limit of the momentum integral is set by the trap depth.  The resulting shift from this effect is then given by:
\begin{equation}
    \label{sm:eq:ahshift}
    \Delta N_{\rm c}^{\mathrm{a.h.}} = N_{\rm c}^{\rm trap} - N_{\rm c}^{\ideal}.
\end{equation}

For the `theoretically expected' value of $\Delta N_{\rm c}^{\rm int}$ we essentially use main text Eq.(1) [see also section \ref{sec:Tccalc}] and the details of how to calculate $S$ are described in section \ref{sec:Scalc}.

\subsection{Data filtering and binning} 

The linear dependence between $N_{\mathrm{th}}$ and $N_0^{2/5}$, that we use for extrapolating to the critical point, only holds for clouds with a BEC present. Also for small values of $N_0$ the $2/5$ scaling means that any noise on $N_0$ can introduce a significant bias. (If the measurements of $N_0$ are represented by a symmetric distribution with the mean corresponding to the true value, the mean value of $N_0^{2/5}$ will always be smaller than the mean value of $N_0$ raised to the power of $2/5$). To counteract these two sources of systematic error we both filter and bin our data before performing a linear fit of $N_{\rm th}$ vs $N_0^{2/5}$.

\begin{figure}
    \phantomsubfloat{\label{sm:fig:N0vst}}
	\phantomsubfloat{\label{sm:fig:Xvst}}
	\centering
    \parbox[c]{\textwidth}{\raggedleft\includegraphics[width=\textwidth]{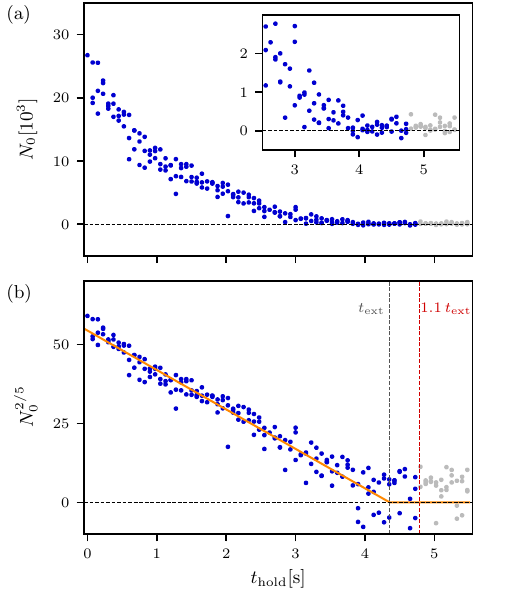}}
	\caption{Procedure for removing purely thermal shots. (a) Time evolution of the BEC number in a typical data series used for extracting the critical point (here $a_s = 72\, a_0$ and $\theta_{\rm d} = 0^{\circ}$). While it is clear that the BEC eventually disappears, it is difficult to directly pinpoint the moment when that happens from the graph; (b) we can linearise the dependence (see text for justification) by plotting $N_0^{2/5}$ against time, as this allows us to define the BEC extinction time using the zero-crossing, $t_{\mathrm{ext}}$, of the fitted truncated-linear trend (orange line). By defining a filter time-window as  $t_{\mathrm{hold}}<1.1 \times t_{\mathrm{ext}}$, we can filter out the points convincingly without any BEC (grey points) and keep the points that (may) contain BEC clouds (blue points).}
	\label{sm:fig:N0filter}
\end{figure}

\textbf{Filtering:} we remove from the dataset all shots for which we can say with certainty, based on hold time, that there is no BEC present. It is difficult to determine the BEC extinction time ($t_{\mathrm{ext}}$) directly from $N_0$ vs $t_{\rm hold}$ [see \cref{sm:fig:N0vst}] so instead in \cref{sm:fig:Xvst} we plot $N_0^{2/5}$ vs $t_{\mathrm{hold}}$ which we expect to be close to linear due to the fact that the thermal atom number, $N_{\mathrm{th}}$, decays approximately linearly in time [see main text Fig.~2(b)]. This allows us to perform a linear fit [see \cref{sm:fig:Xvst}] to determine $t_{\mathrm{ext}}$. In order to capture all the shots that potentially have BECs present, we only remove points that have $t_{\rm hold}> 1.1 \times t_{\mathrm{ext}}$; we note that the final result is insensitive to the exact criterion used.

\textbf{Binning:} To combat potential bias introduced by random noise on $N_0$, we bin and average the data points prior to calculating $N_0^{2/5}$ and performing the fit. The binning is performed over intervals of $t_\textrm{hold}$ that are defined as periods of time over which $N_0$ is expected to change by $\sim 1000$ atoms.
Additionally, we specify a maximum hold time interval of 10\% of the maximum hold time to limit how many points are averaged together.

\section{Theoretical Predictions for $T_{\rm c}$-shift and non-saturation effects }

In this section we: 
\begin{itemize}
    \item Summarize the mean-field $T_{\rm c}$-shift predictions for an anisotropic trap (the result quoted in the main paper is the simplified result for the cylindrically symmetric case).   
    \item Outline how to calculate the Thomas--Fermi radii of a dipolar BEC; these are needed for calculating the non-saturation slopes and are also used in section \ref{sec:Tcorr}.
    \item Calculate the non-saturation effect for our dipolar gas, verifying the linear scaling of $N_{\rm th}$ vs $N_0^{2/5}$ used in our analysis. 
\end{itemize}

\subsection{$T_{\rm c}$-shift predictions in an anisotropic trap}
\label{sec:Tccalc}
In a tri-axial trap the prediction for the temperature shift is~\cite{Glaum:2007b}: 
\begin{equation} \label{sm:eq:TcShiftTh}
    \begin{split}
        \frac{\Delta T_{\rm c}}{T_{\rm c}}  = &
        -3.426  \frac{a_s}{\lambda}\\
        & +3.426 \frac{a_\mathrm{dd}}{\lambda} \frac{1}{2} \Bigg\{ \sin^2{\theta_{\rm d}} \Bigg[ \cos^2{\phi_{\rm d}} f_2{\left(\frac{\omega_x}{\omega_y},\frac{\omega_x}{\omega_z}\right)}\\
        & +\sin^2{\phi_{\rm d}} f_2{\left(\frac{\omega_y}{\omega_z},\frac{\omega_y}{\omega_x}\right)}\Bigg]
        + \cos^2{\theta_{\rm d}} f_2{\left(\frac{\omega_z}{\omega_x},\frac{\omega_z}{\omega_y}\right)}\Bigg\} ~,
    \end{split}
\end{equation}
where the tri-axial anisotropy function modulates the result based on the ratios of the trapping frequencies:
\begin{equation}
    \begin{split}
        f_2{(\eta, \kappa)} =\, &\ 1 + \frac{3 \kappa \eta}{\sqrt{1-\kappa^2}(1-\eta^2)} \Bigg[\\
         & E\left(\sin^{-1}{\sqrt{1-\kappa^2}}, \sqrt{\frac{1-\eta^2}{1-\kappa^2}}\right)\\
         & -F\left(\sin^{-1}{\sqrt{1-\kappa^2}}, \sqrt{\frac{1-\eta^2}{1-\kappa^2}}\right)\Bigg] ~.
    \end{split}
\end{equation}
Here $F(\varphi,k)$ and $E(\varphi,k)$ are the incomplete elliptic integrals of the first and second kind, respectively. Note that the output of $f_2{(\eta, \kappa)}$ is real for all $\eta, \kappa>0$ despite imaginary values appearing in the arguments of the elliptic integrals. 
In a cylindrically symmetric trap, \cref{sm:eq:TcShiftTh} simplifies to \cite{Glaum:2007}:
\begin{equation} \label{sm:eq:Tcshift}
    \frac{\Delta T_{\rm c}}{T_{\rm c}^\ideal} = -3.426  \frac{a_s}{\lambda} + 3.426  \frac{a_{\textrm{dd}}}{\lambda} f{\left(\frac{\omega_{\parallel}}{\omega_{\perp}} \right)} \frac{3 \cos^2{\theta_{\rm d}}-1}{4} ~,
\end{equation}
and the anisotropy function reduces to:
\begin{equation}\label{sm:eq:Cylf}
    f(\kappa) = \frac{2\kappa^2 +1}{1-\kappa^2}-\frac{3 \kappa^2 \tanh^{-1}{\sqrt{1-\kappa^2}}}{(1-\kappa^2)^{3/2}}.
\end{equation}

As we work with close to cylindrically symmetric traps, \cref{sm:eq:Tcshift} still grasps the general behaviour of the system, which is why we quote it, rather than \cref{sm:eq:TcShiftTh}, in the main body of the paper. Nevertheless, we use the full anisotropic version in the data-analysis.

\subsection{Thomas--Fermi solution for dipolar BECs}
\label{sec:TFsol}

Here we outline how to obtain the Thomas--Fermi radii for a dipolar BEC. 
Applying the TF approximation (neglecting the kinetic energy term) to the Gross-Pitaevskii equation gives
\begin{equation}\label{sm:eq:TFeq}
    V{\left(\mathbf{r}\right)} + g n_0{\left(\mathbf{r}\right)} + \Phi_\mathrm{dd}{\left(\mathbf{r}\right)} = \mu 
\end{equation}
where $g=4 \pi \hbar^2 a_s / m$ is the contact interaction strength and $n_0$ the BEC density. The dipolar mean-field potential $\Phi_\mathrm{dd}$ captures the dipole-dipole interactions:
\begin{equation} \label{sm:eq:DDIMF_potential}
\Phi_\mathrm{dd}{\left(\mathbf{r}\right)} = \int{n_0{\left(\mathbf{r}'\right)}
	U_{\mathrm{dd}}{\left(\mathbf{r}'-\mathbf{r}\right)}
	 d\mathbf{r}' } .
\end{equation}

Due to the non-local nature of the DDI, this equation is considerably more complicated to solve compared to the case with purely contact interactions, where famously the solution for the density is an inverted paraboloid:
\begin{equation}\label{sm:eq:TFparabola}
    n_0{\left(\mathbf{r}\right)} = \frac{8 N_0}{15 \pi R_x R_y R_z} \left(1-\frac{x^2}{R_x^2}-\frac{y^2}{R_y^2}-\frac{z^2}{R_z^2} \right),
\end{equation}
with $R_i$ the TF radii of the BEC cloud.

To tackle the DDI case, one can follow the approach from~\cite{Eberlein:2005}, defining the pseudo-potential $\phi{\left(\mathbf{r}\right)}$, satisfying the Poisson equation with the density as the charge term, $\nabla^2\phi{\left(\mathbf{r}\right)}=-n_0{\left(\mathbf{r}\right)}$. This allows for the dipolar mean-field potential to be expressed in the alternative form:
\begin{equation}\label{sm:eq:DDIMF_potential_alt}
    \Phi_\mathrm{dd}{\left(\mathbf{r}\right)} =  -C_\mathrm{dd} \hat{e}_i \hat{e}_j \left(\nabla_i \nabla_j \phi{\left(\mathbf{r}\right)}+\frac{\delta_{ij}}{3}n_0{\left(\mathbf{r}\right)} \right),
\end{equation}
here $C_\mathrm{dd} = \mu_0 \mu_{\rm d}^2 = 12 \pi \hbar^2 a_\mathrm{dd} / m$ is the dipolar interaction strength, $\hat{e}_i$ are the components of the dipole polarisation vector and there is an implicit summation over $i$ and $j$. Expanding $\phi$ as a power series in Cartesian coordinates and considering the symmetry and uniqueness of the solution, using \cref{sm:eq:TFeq,sm:eq:DDIMF_potential_alt} constrains the solution to:  
\begin{equation}
\label{sm:eq:phi}
    \begin{split}
        \phi{\left(\mathbf{r}\right)} = & \phi_0 + \phi_x x^2 + \phi_y y^2 + \phi_z z^2 + \\
        & \phi_{xx} x^4 + \phi_{yy} y^4+ \phi_{zz} z^4 + \\
        & \phi_{xy} x^2 y^2 + \phi_{yz} y^2 z^2+ \phi_{xz} x^2 z^2 ~.                 
    \end{split}
\end{equation}
This implies that the solution for the density can still be written in the form of \cref{sm:eq:TFparabola}, but with the principal axes of the paraboloid potentially tilted w.r.t. to the axes of the trapping potential (the two sets of axes will align if the dipoles are polarised along a trap axis). Switching to this coordinate system and using the fact that we only need to solve for $\phi{\left(\mathbf{r}\right)}$ inside the BEC ellipsoid, the solution takes the form \cite{Eberlein:2005}: 
\begin{equation} \label{sm:eq:phisolTF}
    \begin{split}
        \phi{\left(\mathbf{r}\right)} = & \frac{N_0}{15 \pi} \int_{0}^{\infty}{\Bigg[ 1-} \\
         & 2\left(\frac{x^2}{R_x^2+\sigma}+\frac{y^2}{R_y^2+\sigma}+\frac{z^2}{R_z^2+\sigma} \right) \\
        & +\left(\frac{x^2}{R_x^2+\sigma}+\frac{y^2}{R_y^2+\sigma}+\frac{z^2}{R_z^2+\sigma} \right)^2 \Bigg]\\
        & \frac{d\sigma}{\sqrt{\left(R_x^2+\sigma \right)\left(R_y^2+\sigma \right) \left( R_z^2+ \sigma \right)}} \, ,
    \end{split}
\end{equation}
where $R_i$ are the TF radii along the principal axes of the density ellipsoid. 

The recipe for calculating $\phi{\left(\mathbf{r}\right)}$ also originates from Ref.~\cite{Eberlein:2005}; it relies on the integral:
\begin{equation}
\begin{split}
    \mathcal{I}{\left(R_x, R_y, R_z\right)} & = \\
    & =\int_\tau^\infty{ \frac{d\sigma}{\sqrt{\left(R_x^2+\sigma \right)\left(R_y^2+\sigma \right)\left(R_z^2+\sigma \right)}}} = \\
    &= \frac{2}{\sqrt{R_z^2-R_x^2}}F(\varphi,k)
\end{split}
\end{equation}
where
\begin{equation}
    \varphi = \sin^{-1}{\sqrt{\frac{R_z^2-R_x^2}{R_z^2-\tau}}},
\end{equation}
\begin{equation}
    k = \sqrt{\frac{R_z^2-R_y^2}{R_z^2-R_x^2}},
\end{equation}
and we assume $R_z > R_y > R_x$ (axis relabeling is used where necessary during calculations). With all this in mind, we can express the coefficients in the expansion  [\cref{sm:eq:phi}] of $\phi{\left(\mathbf{r}\right)}$:
\begin{equation}
    \phi_0 = \frac{N_0}{15 \pi}\mathcal{I},
\end{equation}
\begin{equation}
    \phi_i = \frac{4 N_0}{15 \pi}\frac{\partial}{\partial R_i^2}\mathcal{I},
\end{equation}
\begin{equation}
    \phi_{ij} = 
    \begin{cases}
        \frac{4 N_0}{45 \pi}\frac{\partial^2}{\partial^2 R_i^2}\mathcal{I} & i=j, \\
        \frac{8 N_0}{15 \pi}\frac{\partial^2}{\partial R_i^2 \partial R_j^2}\mathcal{I} & i\neq j.
    \end{cases} 
\end{equation}

Using these expressions we numerically find the values of the TF radii, $R_i$, and the spatial orientation of the density ellipsoid, in the trap coordinate frame that constitute a self-consistent solution to the TF equation [\cref{sm:eq:TFeq}]. 

\vspace{0.5cm}
\subsection{Non-saturation of the partially condensed Bose gas}
\label{sec:Scalc}

The non-saturation effect arises due to the ($N_0$-dependent) mean-field interaction potential, resulting from the BEC, that the thermal atoms experience.    
In gases with only contact interactions, this leads to the prediction \cite{Tammuz:2011}: 
\begin{equation}
    N_{\mathrm{th}} = N_{\rm c}^{\ideal}+
    \frac{\zeta{\left( 2 \right)}}{2}
    \left(\frac{k_{\rm B} T}{\hbar \bar{\omega}}\right)^2
    \left(\frac{15 a_s}{\bar{a}_{\mathrm{ho}}}\right)^{\frac{2}{5}}
    N_0^{\frac{2}{5}}
\end{equation}
where $\bar{a}_{\mathrm{ho}} = \sqrt{\hbar / \left( m \bar \omega \right)}$.

With the addition of the dipolar interactions, the effective potential experienced by the thermal gas is:
\begin{equation} \label{sm:eq:MF_totalpotential}
    V_{\mathrm{eff}}{\left( \mathbf{r}\right)}=
    V{\left( \mathbf{r}\right)} 
    + 2 gn_0{\left( \mathbf{r}\right)} + \Phi_\mathrm{dd}{\left(\mathbf{r}\right)}.
\end{equation}
Here we neglect the self-interaction of the thermal gas that is responsible for the interaction $T_{\rm c}$-shift. The dipolar MF potential now takes the form: 
\begin{equation}\label{sm:eq:MF_totalpotential_alt}
    \Phi_\mathrm{dd}{\left(\mathbf{r}\right)} =  -C_\mathrm{dd} \hat{e}_i \hat{e}_j \left(\nabla_i \nabla_j \phi{\left(\mathbf{r}\right)}+2\frac{\delta_{ij}}{3}n_0{\left(\mathbf{r}\right)} \right),
\end{equation}
with the quasi-electrostatic potential again satisfying the Poisson equation $\nabla^2\phi{\left(\mathbf{r}\right)}=-n_0{\left(\mathbf{r}\right)}$. Note that the additional factors of 2 in the \cref{sm:eq:MF_totalpotential,sm:eq:MF_totalpotential_alt} account for the Bose-enhancement of the thermal-BEC atom interactions in a gas of identical bosons.

The quasi-electrostatic potential can be written in the form (similar to \cref{sm:eq:phisolTF}, save for the integration limits):
\begin{equation} \label{sm:eq:phisolNS}
    \begin{split}
        \phi{\left(\mathbf{r}\right)} = & \frac{N_0}{15 \pi} \int_{\tau}^{\infty}{\Bigg[ 1-} \\
          & 2\left(\frac{x^2}{R_x^2+\sigma}+\frac{y^2}{R_y^2+\sigma}+\frac{z^2}{R_z^2+\sigma} \right) \\
        & +\left(\frac{x^2}{R_x^2+\sigma}+\frac{y^2}{R_y^2+\sigma}+\frac{z^2}{R_z^2+\sigma} \right)^2 \Bigg]\\
        & \frac{d\sigma}{\sqrt{\left(R_x^2+\sigma \right)\left(R_y^2+\sigma \right) \left( R_z^2+ \sigma \right)}}.
    \end{split}
\end{equation}
This time, as we are calculating $\phi$ in all space, the integration limit $\tau$ may take different values depending on the region of space---inside the TF ellipsoid we have $\tau = 0$, while outside of the ellipsoid it is found as the (real) root of the equation:
\begin{equation}
    \frac{x^2}{R_x^2 +\tau}+\frac{y^2}{R_y^2 +\tau}+\frac{z^2}{R_z^2 +\tau}=1.
\end{equation}

This allows us to evaluate the MF potential felt by the thermal atoms for any BEC atom number $N_0$. Once the MF potential is calculated, one can perform the same integration as in \cref{sm:eq:psdintegral} to obtain the total number of thermal atoms $N_{\mathrm{th}}$ at a given $N_0$. We numerically confirm that:
\begin{equation}
    \frac{N_{\mathrm{th}}}{N_{\rm c}^{\ideal}} = 1 + S \left(\frac{N_0}{N_{\rm c}^{\ideal}}\right)^{\frac{2}{5}}
\end{equation}
and evaluate $S(a_s,\theta_{\rm d})$ as needed.

\bibliography{robsrefs}